\documentclass[12pt]{article}

\usepackage[dvips,dvipdfm]{graphicx,color}

\usepackage{amsmath,amssymb}
\usepackage{type1cm}
\usepackage{bm}

\numberwithin{equation}{section}

\begin{document}

\begin{titlepage}
\renewcommand{\thefootnote}{\fnsymbol{footnote}}
\vspace*{-5mm}
\hfill
\vbox{
    \halign{#\hfil         \cr
           RIKEN-MP-5 \cr
           } 
      }  
\vspace*{5mm}
\begin{center}
{\Large {\bf 
Toward Bound-State Approach to Strangeness \\ \vspace{3mm} in Holographic QCD
} }

\vspace*{15mm}
{\sc Takaaki Ishii}\footnote{e-mail: {\tt ishiitk@riken.jp}}

\vspace*{7mm} 
{\it Mathematical Physics Laboratory, Nishina Center, RIKEN, 
Saitama 351-0198, Japan
}\\ 

\end{center}

\vspace*{.3cm}
\begin{abstract}
\noindent
An approach to realize a hyperon as a bound-state of a two-flavor baryon and a kaon
is considered in the context of the Sakai-Sugimoto model of holographic QCD,
which approach has been known in the Skyrme model as the bound-state approach to strangeness.
As a simple case of study,
pseudo-scalar kaon is considered as fluctuation around a baryon.
In this case, strongly-bound hyperon-states are absent,
different from the case of the Skyrme model.
Observed is a weak bound-state which would correspond to $\Lambda(1405)$.
\end{abstract}
\vspace{.5cm}

Sep 2010

\pagestyle{empty}

\end{titlepage}

\setcounter{footnote}{0}
\setcounter{tocdepth}{2}


\section{Introduction}

Chiral symmetry plays an important role in low energy of QCD,
where the lightness of pions is because
they could be the Nambu-Goldstone bosons of the spontaneous breaking of the symmetry.
However, it is explicitly broken in \textit{real} QCD by non-zero quark masses.
In particular, the mass of the strange quark should be treated as large
in contrast to those of the up and the down quarks.
The original proposal of the Sakai-Sugimoto model \cite{SaSu1,SaSu2},
which respects the chiral symmetry,
is a holographic dual of massless QCD.
Its quark-mass deformations have also been considered:
A way is to use worldsheet instantons \cite{Hashimoto:2008sr,Aharony:2008an}
(see also \cite{McNees:2008km,Argyres:2008sw,Edalati:2009xc})\footnote{
Another way is a tachyon condensation in the D8/$\overline{\mathrm{D8}}$-branes
\cite{Casero:2007ae,Bergman:2007pm,Dhar:2007bz,Dhar:2008um,Jokela:2009tk}.
}.
In this paper, we focus on baryons with strangeness as a massive flavor in the Sakai-Sugimoto model.

Let us recall that, 
in the context of the Skyrme model,
there is an approach known as the bound-state approach to strangeness \cite{Callan:1985hy,Callan:1987xt}.
It is useful in the case that strangeness is dealt with as a massive flavor.
The idea is to consider a kaon around a two-flavor Skyrmion,
and then their bound-state is identified as a hyperon.
There is a problem that they are somewhat strongly bound such that
hyperon masses are lower than those obtained in experiments,
but the approach seems to work well.

We would like to consider this idea in the Sakai-Sugimoto model,
where baryons are identified as instanton-like solitons \cite{SaSu1}.
The soliton solution of the model has been studied in detail in \cite{HSSY}
(see also \cite{Hong:2007kx,Hong:2007ay,Hong:2007dq,Park:2008sp,HaSaSu1,Kim:2008pw,
Kim:2008iy,HaSaSu2,Kim:2009sr,Hashimoto:2009as}).
We focus on kaon fluctuation here as a simple attempt to the bound-state approach.
This attempt should be different from the collective coordinate quantization of three flavor baryons \cite{HataMurata},
in which case mass shifts of baryon spectra in the presence of quark masses have been calculated in \cite{Hashimoto:2009st}\footnote{
Baryon mass shifts in the case of two flavors have been studied in \cite{Hashimoto:2009hj}.
}.
Our study may rather be related to meson-baryon scatterings such as $\bar{K}$-$N$ scatterings.

This paper is organized as follows.
In Section \ref{sec:kaon}, we consider kaon fluctuation around the baryon as the background of the model.
Here for simplicity we try to use a canonical kaon mass term.
In Section \ref{sec:mass}, we consider the quark-mass term introduced by \cite{Hashimoto:2008sr,Aharony:2008an}.
This mass term should be a proper one to be used instead of the one used in Section \ref{sec:kaon},
while we will see that its contribution is minor here.
In Section \ref{sec:summary}, we summarize, and
briefly discuss a possibility to take vector mesons into account.

\section{Kaon fluctuation around a baryon}
\label{sec:kaon}

\subsubsection*{Summary of the Sakai-Sugimoto model}

The Sakai-Sugimoto model \cite{SaSu1,SaSu2} is a five-dimensional
gauge theory, in which
Kaluza-Klein(KK)-decomposed gauge fields are mesons and baryons are provided by its solitons.

The action is given by the following five-dimensional $U(N_f)$ Yang-Mills-Chern-Simons theory 
in a curved background:
\begin{eqnarray}
&&
S_\mathrm{SS} =
-\kappa \int d^4 x dz \,\mathrm{Tr}
\left[
\frac{1}{2} h(z) F_{\mu\nu}^2 + M_\mathrm{KK}^2 k(z) F_{\mu z}^2
\right]
+ \frac{N_c}{24\pi^2} \int_\mathrm{M_5} \omega_5.
\label{SS_action}
\end{eqnarray}
The functions $h(z) = (1+z^2)^{-1/3}$ and $k(z) = 1+z^2$ inherit the metric of the curved background.
The field strength is given by $F = d A + i A \wedge A$ with the $U(N_f)$ flavor gauge fields $A_\mu$ and $A_z$, and
$\omega_5 = \mathrm{Tr} \left( A \, F^2 - \frac{i}{2}A^3 F - \frac{1}{10} A^5 \right)$
is the Chern-Simons five-form.
This model describes a holographic dual of a large $N_c$ {\it massless} QCD.

There are two parameters in \eqref{SS_action}: $M_\mathrm{KK}$ and $\kappa$.
For the mass scale $M_\mathrm{KK}$, we chose a unit in which $M_\mathrm{KK} = 1$.
The dimensionful parameter can be easily recovered from dimensional analysis.
The other is given by $\kappa = \lambda N_c/(216\pi^3)$ with the 't~Hooft coupling constant $\lambda$.
For later convenience we define $a \equiv 1/(216\pi^3)$ such that $\kappa = a \lambda N_c$.
These are chosen as
\begin{eqnarray}
M_\mathrm{KK} = 949 \, \mathrm{[MeV]}, \quad  \kappa = 0.00745
\label{values_of_M_KK_and_kappa},
\end{eqnarray}
to fit the experimental values\footnote{
If one rigorously treats the masslessness,
numerical values in the chiral limit should be used for $f_\pi$ and $m_\rho$.
}
of the $\rho$ meson mass $m_\rho \simeq 776$ MeV 
and the pion decay constant $f_\pi \simeq 92.4$ MeV.
We use \eqref{values_of_M_KK_and_kappa} in this paper\footnote{
One may fit the parameters using baryon data
instead of using $m_\rho$.
If the mass difference of $\Delta$ and $N$ is used,
$M_\mathrm{KK} \sim 500$ MeV \cite{HSSY},
and then $\kappa$ can be calculated accordingly from $f_\pi$ and $M_\mathrm{KK}$.
}.

A baryon is identified as an instanton-like soliton localized in the
four-dimensional space $x^\alpha$ ($\alpha=1,2,3,z$) \cite{SaSu1}.
The instanton number corresponds to the baryon number. 
Let us see the case that $N_f=2$.
A solution of the equations of motion of \eqref{SS_action} 
has been obtained in \cite{HSSY} in the leading order in $1/\lambda$.
The spatial component $A_\alpha^\mathrm{inst}$ is precisely the $SU(2)$ BPST instanton,
and also there is a nonzero $U(1)$ part $A_0^\mathrm{inst}$;
in singular gauge, they are given by
\begin{eqnarray}
&&A_i^{\mathrm{inst}} =
-\frac{\rho^2}{\xi^2(\xi^2+\rho^2)}
((z-Z) \tau_i + \epsilon_{ijk} (x_j-X_j) \tau_k),
\quad
A_z^{\mathrm{inst}} =
\frac{\rho^2}{\xi^2(\xi^2+\rho^2)} (x_i-X_i) \tau_i,
\nonumber \\
&&A_0^\mathrm{inst} =
\frac{1}{16 \pi^2 a \lambda}
\frac{1}{\xi^2}
\left[
1-\frac{\rho^4}{(\rho^2+\xi^2)^2}
\right]
\mathbf{1}_2 ,
\label{SU2_instanton_solution}
\end{eqnarray}
where $\xi \equiv ( (z-Z)^2 + |\vec{x} - \vec{X}|^2 )^{1/2}$ is the distance in the spatial four dimensions,
and $\tau_i \ (i=1,2,3)$ are the Pauli matrices.
The location of the soliton is $X^\alpha = (X^1,X^2,X^3,Z) = (\vec{X},Z)$, 
and $\rho$ is the instanton size.
These correspond to five of the eight moduli of a $SU(2)$ instanton.
Classically, the baryon is at the bottom of the curved space and has a finite size,
\begin{eqnarray}
Z_\mathrm{cl}=0,
\quad
\rho_\mathrm{cl}^2
= \frac{1}{8\pi^2a\lambda} \sqrt{\frac{6}{5}}.
\label{baryon_classical_rho_z}
\end{eqnarray}
The non-zero value of $\rho_\mathrm{cl}$ is thanks to $A_0^\mathrm{inst}$,
which prevents the instanton shrinks to zero-size in the curved space.
Thus $A_0^\mathrm{inst}$ is qualitatively important.
In the following the baryon is placed at $\vec{X}=0$.

\subsubsection*{Kaon fluctuation in the baryon background}

Let us consider kaon fluctuation around the two-flavor baryon.
Our strategy is such that
we first reduce the five-dimensional action \eqref{SS_action} to four-dimensional one
which representing a kaon in a potential,
and then consider bound-state problem following \cite{Callan:1985hy,Callan:1987xt}.
In the following, we set $N_f=3$.

The kaon can be turned on as a fluctuation of gauge fields around
the two-flavor baryon as a instanton-like background gauge field.
The two-flavor configuration \eqref{SU2_instanton_solution} is embedded in $3 \times 3$ matrices,
and fluctuation $a_z$, which is a two-component vector,  is turned on at off-diagonal components,
\begin{eqnarray}
A_\mu =
\begin{pmatrix}
A_\mu^{\mathrm{inst}} & 0 \\  0 & 0
\end{pmatrix},
\quad
A_z =
\begin{pmatrix}
A_z^{\mathrm{inst}} & a_z \\ a_z^\dagger & 0
\end{pmatrix}.
\label{SU3_strange_kaon_fluctuation}
\end{eqnarray}
We assume that,
even in the presence of the baryon as the background,
$a_z$ is decomposed into the four-dimensional kaon field as
\begin{eqnarray}
a_z = -\frac{1}{\sqrt{2}} K(x^\mu) \phi_0(z),
\quad
 K(x^\mu) = 
 \begin{pmatrix}
K^+ \\ K^0
\end{pmatrix},
\label{KK_kaon_zero_mode}
\end{eqnarray}
where we are using the gauge in which the KK-modes with $n \ge 1$ are gauged away.
The factor in front of $K(x^\mu)$ corresponds to the canonical normalization of $K(x^\mu)$.
The zero-mode $\phi_0$ is chosen such that
\begin{eqnarray}
\phi_0(z) = \frac{1}{\sqrt{\kappa \pi}} \frac{1}{k(z)},
\quad
\kappa \int_{-\infty}^{\infty} dz \, k(z) \phi_0(z)^2 = 1.
\label{phi0_zeromode}
\end{eqnarray}
To be precise, \eqref{KK_kaon_zero_mode} and \eqref{phi0_zeromode} were obtained
when quarks are massless and baryons are absent \cite{SaSu1}.
We believe that the reduction to four dimensions by using these
would be a convenient assumption and approximation.

A four-dimensional action of $K(x^\mu)$ gives a starting point of our bound-state approach.
Putting \eqref{SU3_strange_kaon_fluctuation} into \eqref{SS_action},
and keeping the terms quadratic in $a_z$, we obtain
\begin{eqnarray}
S_\mathrm{fluc}
=
-2 \kappa \int d^4 x dz \,
k(z)
\bar{D}_\mu^\mathrm{(inst)} a_z^\dagger D^{\mathrm{(inst)} \mu} a_z,
\label{5dim_kaon_action}
\end{eqnarray}
where the covariant derivatives are given by
\begin{eqnarray}
D_\mu^\mathrm{(inst)} a_z
= \partial_\mu a_z + i A_\mu^\mathrm{inst} a_z,
\quad
\bar{D}_\mu^\mathrm{(inst)} a_z^\dagger
= \partial_\mu a_z^\dagger 
-i a_z^\dagger A_\mu^\mathrm{inst}.
\label{covariant_derivative_az}
\end{eqnarray}
To do the integral over $z$, we use \eqref{phi0_zeromode}, and then obtain
\begin{eqnarray}
S_\mathrm{kaon}
= - \int d^4x \left[
\partial_\mu K^\dagger \partial^\mu K
+i \left(
\partial^\mu K^\dagger\, \Psi_\mu K
- K^\dagger\, \Psi_\mu \partial^\mu K
\right)
+ K^\dagger \Upsilon K 
\right],
\label{kaon_fluctuation_4dim_action}
\end{eqnarray}
where
\begin{eqnarray}
\Psi_\mu \equiv \frac{1}{\pi}\int dz \frac{1}{k(z)} A_\mu^\mathrm{inst},
\quad
\Upsilon \equiv \frac{1}{\pi}\int dz \frac{1}{k(z)}
(A_\mu^\mathrm{inst})^2.
\end{eqnarray}
Note that $\Psi_i$ is written in Pauli matrices, while $\Psi_0$ and $\Upsilon$ are 
proportional to unit matrices,
\begin{eqnarray}
&& 
\Psi_i =
- \Theta^{(1)}
\epsilon_{ijk}x_j\tau_k, \quad
\Psi_0 = 
\frac{1}{16\pi^2 a\lambda}
\Theta^{(2)}\mathbf{1}_2,
\nonumber \\
&& 
\Upsilon
= \left[ 3 \Theta^{(3)} + 2 r^2 \Theta^{(4)} - \left( \frac{1}{16\pi^2 a\lambda}\right)^2 \Theta^{(5)} \right] \mathbf{1}_2,
\end{eqnarray}
where $\Theta^{(n)}$ are given in Appendix.
It is convenient to define
$\Psi_0 \equiv \widehat{\Psi}_0 \mathbf{1}_2$ and 
$\Upsilon \equiv \widehat{\Upsilon} \mathbf{1}_2$.

Note that \eqref{SU2_instanton_solution} is a flat-space instanton.
This is obtained in $z \ll 1$ region, where the space is almost flat $k(z), h(z) \sim 1$.
Effects of the curved space come in at large $z$ region.
However, in the present case we are interested in the overlap of $\phi_0$ and the baryon,
both of which are localized at $z=0$.
To use \eqref{SU2_instanton_solution} would be a meaningful approximation.

\subsubsection*{Equation of motion}

Recall that \eqref{kaon_fluctuation_4dim_action} corresponds to massless kaon.
We therefore need a mass term.
In this section we try to use the canonical kaon mass term for simplicity,
\begin{eqnarray}
S_\mathrm{kaon} - \int d^4 x \, m_K^2 K^\dagger K,
\end{eqnarray}
where $m_K \simeq 495$ MeV.
We will consider a quark mass term which is precisely introduced to the Sakai-Sugimoto model
separately in the next section.

Since we have the four-dimensional action \eqref{kaon_fluctuation_4dim_action},
we may simply trace the analysis of \cite{Callan:1985hy,Callan:1987xt}.
We decompose the kaon field by creation and annihilation operators as\footnote{
The signs of the exponents are different from \cite{Callan:1985hy,Callan:1987xt}
due to our convention.
}
\begin{eqnarray}
K(x^\mu) = \sum_{n>0}
\left( k_n(r) Y_{T T_z L}(\Omega_2) e^{i \omega_n t} a_n + \tilde{k}_n(r) Y_{T T_z L}(\Omega_2) e^{-i \tilde{\omega}_n t} b_n^\dagger \right),
\end{eqnarray}
where $Y_{T T_z L}$ is spherical harmonics on $S^2$.
The basis of the harmonics is labeled by angular momentum $\bm{L}$ and total spin $\bm{T} = \bm{I} + \bm{L}$,
where $\bm{I} = \bm{\tau}/2$.

After a short calculation, we obtain the following eigenvalue equations,
\begin{eqnarray}
&& \left[ \frac{1}{r^2} \partial_r \left( r^2 \partial_r \right) + \omega_n^2 + 2 \widehat{\Psi}_0 \, \omega_n - V(r) \right] k_n(r) = 0,
\label{eigenvalue_eom}  \\
&& \left[ \frac{1}{r^2} \partial_r \left( r^2 \partial_r \right) + \tilde{\omega}_n^2 - 2 \widehat{\Psi}_0 \, \tilde{\omega}_n- V(r) \right] \tilde{k}_n(r) = 0,
\label{exotic_eom}
\end{eqnarray}
where
\begin{eqnarray}
V(r) = \frac{L(L+1)}{r^2} + 4 \Lambda_{\bm{I \cdot L}}
 \Theta^{(1)} + \widehat{\Upsilon} + m_K^2,
 \label{bound_state_potential}
\end{eqnarray}
with the eigenvalue of the spin-isospin interaction given by
\begin{eqnarray}
\bm{I \cdot L} \, Y_{T T_z L}(\Omega_2)
= \frac{T(T+1) - L(L+1) - 3/4}{2} \, Y_{T T_z L}(\Omega_2)
\equiv \Lambda_{\bm{I \cdot L}} \, Y_{T T_z L}(\Omega_2).
\end{eqnarray}
As seen from the signs in front of $\widehat{\Psi}_0$,
\eqref{eigenvalue_eom}  and \eqref{exotic_eom} correspond to 
$S=-1$ (normal) and $S=+1$ (exotic) channels, respectively.
Below we focus on \eqref{eigenvalue_eom}.
In the Skyrme model, the exotic channel was studied in the context of penta-quarks \cite{Itzhaki:2003nr}.
Note that, these equations are somehow different from those obtained in \cite{Callan:1985hy,Callan:1987xt}.
Firstly, since the kinetic term of \eqref{kaon_fluctuation_4dim_action} is canonical,
no additional functions of $r$ appear in $\partial_r$ and $\omega$ terms.
Secondly, we will see that the behavior of $V(r)$ is repulsive to dislike bound-states.

We shall consider two cases of $(T,L)$: $T=\frac{1}{2}, L=1$ and $T=\frac{1}{2}, L=0$.
In \cite{Callan:1985hy,Callan:1987xt},
the former case corresponds to $\Lambda$, $\Sigma$ and $\Sigma^\ast$, 
while the latter case did $\Lambda(1405)$;
for details, see the papers.
Such identification was because the bound-states could be found,
and the spectra were nicely interpreted as those baryons.
Relying on the observations, we now search whether there are bound-states in such two cases.

Solving the eigenvalue problems, we find no bound-state in the $T=\frac{1}{2}, L=1$ case,
while a weak bound-state can be found in the $T=\frac{1}{2}, L=0$ case with $\omega_1 \sim 490$ MeV.
In the former case, $V(r)$ is strongly repulsive opposite to the Skyrme model case,
where the potential are strongly attractive and favor a bound-state.
In the latter case, the potential is still repulsive (see Figure \ref{fig:pot}).
However, in practice $\Psi_0$ in \eqref{eigenvalue_eom} gives seemingly large contribution
which forces kaon wavefunction localize at the slight minimum of $V(r)$ in this case.
We concern that this may be a numerical artifact of the baryon solution of the Sakai-Sugimoto model
because $A_0^\mathrm{inst}$,
which should have been of order $\frac{1}{\lambda}$ and small, 
is in fact large due to the numerical factor appearing in it.

It seems that our bound-state program tries to bind a kaon and a baryon,
both of which have concrete identity after that they are made from a quark-antiquark pair or $N_c$ quarks.
This point may be related to a prescription where $\Lambda(1405)$ can be a bound-state of $\Bar{K}$ and $N$.

We do not step in the $SU(2)$ corrective coordinate quantization of the baryon \cite{HSSY}
although the calculation is straightforward.
It will produce the mass splittings of $\Lambda$, $\Sigma$ and $\Sigma^\ast$ baryons, for example.
Since we could not obtain any bound-state corresponding in $T=\frac{1}{2}, L=1$ case,
such mass splittings cannot be discussed.

\section{Kaon mass term revisited}
\label{sec:mass}

In the previous section, the canonical kaon mass term $m_K^2 K^\dagger K$ was employed.
However, there are precise ways to introduce quark masses by using ingredients of string theory.
In this section, we consider the quark mass term introduced in \cite{Hashimoto:2008sr}.
We will see that its contribution is minor in the present case of the bound-state problem.
For this reason we used $m_K^2 K^\dagger K$ phenomenologically in the previous section.

The quark-mass term introduced in \cite{Hashimoto:2008sr} is
\begin{eqnarray}
 \delta S = c \int d^4x \; \mathrm{PTr} 
\left[ M \left(
\exp \left[-i \int_{-z_m}^{z_m} A_z dz \right]
- \mathbf{1}_{N_f} \right) \right] + \mathrm{c.c.}
\label{worldsheet_boundary_coupling}
\end{eqnarray}
This term can be induced from worldsheet instanton amplitude 
in the case that D6-branes are put into the D-brane configuration of the Sakai-Sugimoto model.
Here $c$ is a constant which relates quark masses to meson masses.
We leave it undetermined, but we will soon rewrite the mass term in terms of the meson masses.
Because \eqref{worldsheet_boundary_coupling} is introduced as perturbation from the chiral limit,
$z_m$ must be close enough to $z=\infty$,
and hence we may approximate the integral region as
\begin{eqnarray}
 \delta S = c \int d^4x \; \mathrm{PTr} 
\left[ M \left(
\exp \left[-i \int_{-\infty}^{\infty} A_z dz \right]
+ \exp \left[i \int_{-\infty}^{\infty} A_z dz \right]
- 2 \, \mathbf{1}_{N_f} \right) \right].
\label{additional_action}
\end{eqnarray}
For simplicity, we set $m_u=m_d=m_0$ such that $M = \mathrm{diag}(m_0,m_0,m_s)$.
Below we derive a kaon mass term in the baryon background.

The path-ordering with respect to the background $A_z^\mathrm{inst}$ is abelian.
This is because $A_z^\mathrm{inst}$ is proportional only to a matrix $\vec{x}\cdot \vec{\tau}$ \cite{Atiyah:1989dq}.
Thus the path-ordering reduces to an ordinary integral,
\begin{eqnarray}
\mathrm{P}\exp \left[-i \int_{a}^{b} dz
\begin{pmatrix}
A_z^\mathrm{inst} & 0 \\ 0 & 0
\end{pmatrix}
\right] = 
\begin{pmatrix}
\exp \left[-i \int_{a}^{b} A_z^\mathrm{inst} dz \right] & 0 \\ 0 & 1
\end{pmatrix}.
\end{eqnarray}
For convenience, we use a notation
\begin{eqnarray}
\int_a^b dz A_z^\mathrm{inst} = I_{(a,b)} \hat{x}^i \tau_i,
\end{eqnarray}
where $\hat{x}_i = x_i/r$ are unit vectors $(r = |\vec{x}-\vec{X}|)$, and
\begin{eqnarray}
I_{(a,b)} =
\left[
\tan^{-1} \left( \frac{b}{r} \right) -\frac{r}{\sqrt{r^2+\rho^2}} \tan^{-1} \left( \frac{b}{\sqrt{r^2+\rho^2}} \right)
\right]
- \left( b \to a \right).
\end{eqnarray}

However, the path-ordering is no longer abelian if $a_z$ is turned on.
Putting \eqref{SU3_strange_kaon_fluctuation} in \eqref{additional_action} and keeping terms of order $a_z^\dagger a_z$, 
we obtain
\begin{eqnarray}
S_\mathrm{mass}
&=& - \int d^4x
\left[
f_\pi^2 \left( m_K^2 - \frac{m_\pi^2}{2} \right)
\int_{-\infty}^{\infty} dz_2  \int_{-\infty}^{z_2} dz_1
\cos (I_{(z_1,z_2)}) a_z^\dagger(z_1) a_z(z_2)
\right. \nonumber \\ 
&& 
\left. + \frac{f_\pi^2 m_\pi^2}{2}
\int_{-\infty}^{\infty} dz_1 \int_{-\infty}^{z_1} dz_2
\cos ( I_{(-\infty,\infty)} - I_{(z_2,z_1)} ) a_z^\dagger(z_2) a_z(z_1) \right],
\label{nonlocal_5dim_kaon_mass_term}
\end{eqnarray}
where the constant $c$ and the quark masses are rewritten in terms of the meson masses \cite{Hashimoto:2009st},
\begin{eqnarray}
m_\pi^2 = \frac{4c}{f_\pi^2} m_0,\quad 
m_K^2 = \frac{2c}{f_\pi^2} (m_0+m_s).
\label{3_flavor_quark_mass_meson_mass}
\end{eqnarray}
The path-ordering is almost abelian where two fluctuations $a_z$ and $a_z^\dagger$ are inserted.
The double-integrals in \eqref{nonlocal_5dim_kaon_mass_term} are with respect to the positions of them.

\begin{figure}
\centering
\includegraphics[width=6cm,clip]{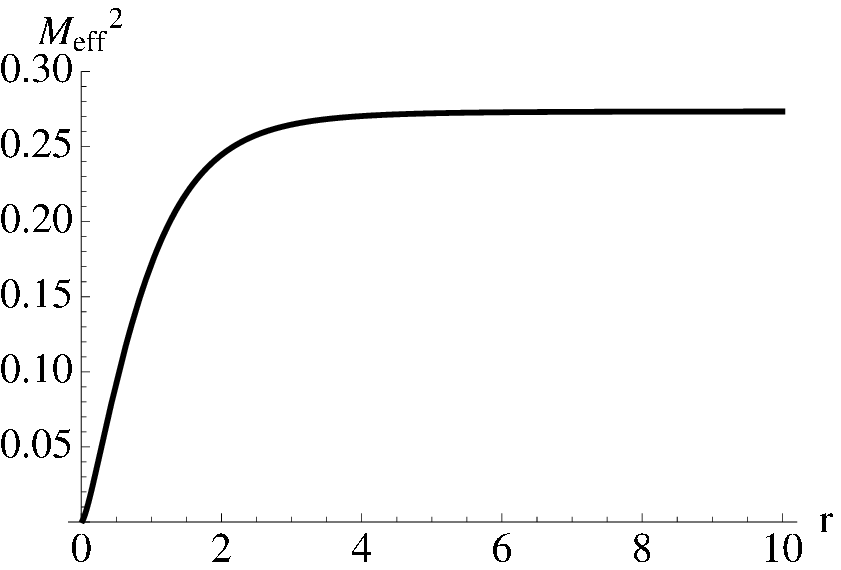}
\caption{Behavior of $M_\mathrm{eff}^2(r)$. $M_\mathrm{eff}=m_K$ at $r=\infty$ while $M_\mathrm{eff}=0$ at $r=0$.}
\label{fig:mk}
\end{figure}

We can obtain an effective kaon mass term
after reducing \eqref{nonlocal_5dim_kaon_mass_term} to four dimensions.
We have
\begin{eqnarray}
S_\mathrm{mass} =
- \int d^4x \, M_\mathrm{eff}^2(r) K^\dagger(x^\mu) K(x^\mu),
\label{effective_kaon_mass_term}
\end{eqnarray}
where the effective kaon mass of is a function of $r$,
\begin{eqnarray}
M_\mathrm{eff}^2(r) &=& 
\frac{2}{\pi^2} 
 \int_{-\infty}^{\infty} dz_2  \int_{-\infty}^{z_2} dz_1\frac{1}{k(z_1)k(z_2)}
\nonumber \\
&& \times
\left[ \left( m_K^2 - \frac{m_\pi^2}{2} \right) \cos ( I_{(z_1,z_2)} ) 
+ \frac{m_\pi^2}{2}
\cos ( I_{(-\infty,\infty)} - I_{(z_1,z_2)} ) \right].
\label{effective_kaon_mass}
\end{eqnarray}
The behavior of $M_\mathrm{eff}$ is plotted in Figure \ref{fig:mk}.
Here we set $m_0=0$ for simplicity.
$M_\mathrm{eff}$ goes to $m_K$ away from the baryon,
while it vanishes sufficiently at the center of the baryon.
Note that the baryon is now given by a finite-size soliton.

\begin{figure}
\centering
\includegraphics[width=6cm,clip]{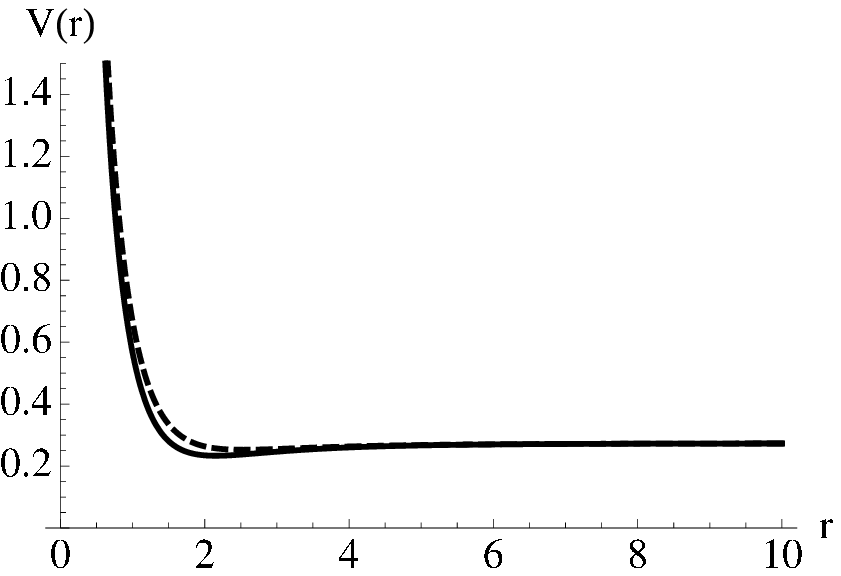}
\caption{$V(r)$ (dashed), and its counterpart if $M_\mathrm{eff}$ is used instead of $m_K$ (solid).}
\label{fig:pot}
\end{figure}

In the case that \eqref{effective_kaon_mass} is used instead of $m_K^2$,
the potential \eqref{bound_state_potential} is slightly modified as shown in Figure \ref{fig:pot}.
The effect is minor since the region where $M_\mathrm{eff}^2$ drops is where $V(r)$ grows up\footnote{
If some contribution, 
as will be discussed in section \ref{sec:summary}, 
drastically changes the repulsive behavior of the potential,
then the effect of the reduce of $M_\mathrm{eff}^2$ at small $r$ will be significant.
}.
The bound-state energy in the $T=\frac{1}{2}, L=0$ case is not largely modified:
It decrease only a few MeV.
There is still no bound-state in the $T=\frac{1}{2}, L=1$ case.

\section{Summary and discussion}
\label{sec:summary}

Toward the bound-state approach to strangeness in the context of the Sakai-Sugimoto model of holographic QCD,
we considered, as a simple case of study, fluctuation which corresponds to kaon around the instanton-like baryon.
We then used a four-dimensional action reduced from five-dimensional one.
Suggested by analyses in the Skyrme model \cite{Callan:1985hy,Callan:1987xt},
we focused on two cases which characterized the kaon wavefunction:
the $T=\frac{1}{2}, L=1$ case and the $T=\frac{1}{2}, L=0$ case.
In both cases, the potential which the kaon feels were repulsive.
There was no bound-state in the former case,
while a weak bound-state was seen in the latter case
although this might be a numerical artifact of the baryon solution of the Sakai-Sugimoto model.
This would correspond to a $\Lambda(1405)$ described by a meson-baryon bound-state.

The disparity of ours from the case of the Skyrme model is in a sense apparent.
In the latter case, the kaon fluctuation is smoothly connected to 
the $SU(3)$ collective coordinate quantization of the Skyrmion in the chiral limit.
However, as long as focusing on kaon fluctuation as in \eqref{SU3_strange_kaon_fluctuation},
the collective coordinate quantization of the three-flavor baryon \cite{HataMurata} should not be realized
because all gauge field fluctuations need to be included so as to generate the instanton zero-mode.

If we want to find connections to collective coordinate quantization description,
it may be necessary to include also vector-mesons.
In fact, a work of bound-state approach in hidden local symmetry framework \cite{Scoccola:1989yk},
which work contained a vector-meson counterpart of kaon,
could give a result similar to the case of the Skyrme model.
In this line, there is another work
which considered vector-mesons of pions but that of the kaon is absent
\cite{Scoccola:1988wa}.
The latter may be similar to our attempt.

However, dealing with vector-mesons in the case here will be far more difficult for following reasons.
First, quark masses should be included in a way that 
mass spectra of vector-mesons can be calculated in the presence of them.
Secondly, effective vector meson masses should be calculated in baryon background as in Section \ref{sec:mass}.
To obtain an effective low-energy description
which will be in the same universality class of the bound-state approach in the Skyrme model,
all the vector mesons (of kaons) should be integrated out at some energy scale.
It is a question whether this story really works.

There is a gauge so-called ``$A_z=0$ gauge'' in the Sakai-Sugimoto model.
In this gauge, the Skyrme model action can be obtained from the Sakai-Sugimoto model;
pion field appears from non-normalizable mode at the boundary of the curved space.
But baryons are located at the bottom of it.
Hence, to study meson-baryon systems we should know the overlap of these two.
Using this gauge may be in direct connection to the bound-state approach in the Skyrme model,
but whether this is the case or not remains to be seen.

\section*{Acknowledgments}

The author would like to thank
Minoru Eto, Koji Hashimoto, Hideaki Iida, Akitsugu Miwa and Sanefumi Moriyama
for fruitful discussions and comments.
The author is supported by JSPS Research Fellowships for Young Scientists.

\appendix

\section{Appendix}
The following symbols are used for convenience:
\begin{eqnarray}
\Theta^{(1)} &\equiv& \frac{1}{\pi} \int_{-\infty}^{\infty} dz \frac{1}{1+z^2} \frac{\rho^2}{\xi^2(\xi^2+\rho^2)}, \\
\Theta^{(2)} &\equiv& \frac{1}{\pi} \int_{-\infty}^{\infty} dz \frac{1}{1+z^2} \frac{1}{\xi^2} \left( 1-\frac{\rho^4}{(\rho^2+\xi^2)^2} \right), \\
\Theta^{(3)} &\equiv& \frac{1}{\pi} \int_{-\infty}^{\infty} dz \frac{1}{1+z^2} (z-Z)^2 \left( \frac{\rho^2}{\xi^2(\xi^2+\rho^2)} \right)^2, \\
\Theta^{(4)} &\equiv& \frac{1}{\pi} \int_{-\infty}^{\infty} dz \frac{1}{1+z^2} \left( \frac{\rho^2}{\xi^2(\xi^2+\rho^2)} \right)^2, \\
\Theta^{(5)} &\equiv& \frac{1}{\pi} \int_{-\infty}^{\infty} dz \frac{1}{1+z^2} \left[ \frac{1}{\xi^2} \left( 1-\frac{\rho^4}{(\rho^2+\xi^2)^2} \right) \right]^2.
\end{eqnarray}

\providecommand{\href}[2]{#2}\begingroup\raggedright\endgroup

\end{document}